\documentclass{article}
\usepackage{spconf,amsmath,graphicx}

\usepackage{algorithm}
\usepackage{algorithmic}

\usepackage{subfig}

\title{A WAVELET-BASED APPROACH TO MONITORING PARKINSON'S DISEASE SYMPTOMS}

\name{Avishai Wagner \qquad Naama Fixler \qquad Yehezkel S. Resheff}

\address{Intel Advanced Analytics}

\begin{document}

\maketitle

\begin{abstract}
Parkinson's disease is a neuro-degenerative disorder affecting tens of millions of people worldwide. Lately, there has been  considerable interest in systems for at-home monitoring of patients, using wearable devices which contain inertial measurement units. We present a new wavelet-based approach for analysis of data from single wrist-worn smart-watches, and show high detection performance for tremor, bradykinesia, and dyskinesia, which have been the major targets for monitoring in this context. We also discuss the implication of our controlled-experiment results for uncontrolled home monitoring of freely behaving patients. 
\end{abstract}

\begin{keywords}
Parkinson's Disease, Monitoring, Wearable Devices, Accelerometer, Wavelets
\end{keywords}

\section{Introduction}
\label{sec:intro}

Parkinson's Disease (PD) is a neuro-degenerative disorder affecting tens of millions of people worldwide \cite{vos2015global}. PD is associated with motor symptoms such as tremor, rigidity, bradykinesia and gait impairments. Another motor difficulty is Levodopa induced dyskinesia -- a side effect of chronic Levodopa therapy, characterized by involuntary and unorganized movement. Currently, the accepted clinical measurement of PD symptom severity is the Unified Parkinson’s Disease Rating Scale (UPDRS), which is based, in part, on subjective and potentially recall-distorted reports by the patients, and on semi-objective observations by the clinician \cite{goetz2008movement}. The UPDRS is typically assessed merely two to three times a year (when the clinician meets the patient face to face), and therefore tends to miss occasional fluctuations in the state of the patients. These fluctuations are a very common phenomenon in PD. The lack of measure continuity and objectivity allows for variation and potential biases in diagnoses or determinations of PD-related motor states. This may adversely affect the prescribed treatment plan.

Partially due to the limitations of the UPDRS standard, PD has been an early target for experimentation with monitoring using wearable devices which contain inertial measurement units \cite{maetzler2013quantitative}. Replacing subjective patient self-reports with objective measures may improve and standardize the way physicians understand the actual state of the disease. In addition, long-term monitoring of symptoms will allow a system to detect and flag changes requiring intervention, thus improving the care patients are able to receive. Finally, longitudinal continuous collection of clinical objective measures from a large cohort can reduce clinical trial costs and duration, and accelerate research in the field of PD.

There have been several attempts to design systems for PD monitoring with wearable devices. MercuryLive \cite{chen2011web, Patel2010} is a system for at-home monitoring of PD patients, using a network of SHIMER sensors \cite{o2009shimmer} to monitor symptoms and activity. Kinesia system \cite{giuffrida2009clinically} is comprised of a single 12 gram finger worn sensor and an 85 gram wrist worn computing module, connected via a wire. The predicted tremor score, output by this system, was shown to correlate with clinical scores \cite{mostile2010correlation}. Although demonstrating the ability to extract some significant objective measures from wearable sensors, the equipment used in these systems poses significant scalability limitations, due to its cost, complexity and burden to the patient. 

An efficient and usable monitoring system should include a single wearable element \cite{espay2016technology}; if possible, a standard unobtrusive device. Few researchers explored the potential of using a consumer smartphone as a measurement device for PD based on a set of predefined tests \cite{arora2015detecting}. This approach requires a significant commitment from the patient, and doesn't necessarily reflect natural activities of daily living.

Previously, analysis of accelerometer signals for symptom prediction \cite{patel2007analysis} or behavior detection \cite{resheff2014accelerater} has mostly been done using hand-crafted features, suitable for the task at hand, and demonstrated on data collected in highly controlled environments. The main contribution of this paper is in the novel wavelet-based feature engineering approach, in the context of PD patient monitoring. We show that our method leads to high accuracy classification of multiple PD-related symptoms, using data from single wrist-worn smart-watches. Our method is appropriate for long term at-home monitoring of freely behaving individuals, as a long-term, unobtrusive, wearable solution.

\section{Data Collection}
\label{sec:data-coll}

Data was collected from 19 PD patients  (age: $62 \pm 8.8$ years; disease duration: $8 \pm 4.5$ years), during two days of hospital visits, and two additional days of home monitoring. During each visit the patient preformed 20 motor tasks. Each task was repeated 6 times in each visit, resulting in 240 task segments per patient. 

Motor tasks performed by the patients were divided into 5 general groups of activity: Resting (e.g. sitting quietly), gross upper limb movement (e.g. folding towels), fine upper limb (e.g. drawing), periodic hand movement (e.g. hand rotation), and walking. 

During each task, a physician graded a number of symptoms (tremor, dyskinesia, bradykinesia) on a score from 0 to 4, where 0 indicates the symptom is not present, and 4 indicates a strong presence. For modeling purposes, we grouped scores 2, 3 and 4 into one bin, due to imbalanced distribution of the values. We treat the labels as ground truth while being aware of the inter-rater variability associated with such tests, and its implications on the quality of automatic classification.  

Subjects arrived at the clinic twice, in different stages of the medication effect cycle, to allow observation of the same patients with differential symptom severity. The first visit started in the “on” state during which the patient experienced a positive response to medication, and continued through the transitioning to the “off” state, during which the patient experienced a reemergence of the Parkinsonian symptoms suppressed during the "on" state. The second visit started in the clinical “off” state, following at least 12 hours without medication intake, and following medication intake, continued through the transitioning of the subject to the “on” state.

Raw tri-axial accelerometer data was collected using a single wrist-worn smart-watch (GENEActiv\footnote{www.geneactiv.org}), at 50Hz throughout the visit, including the full duration of the motor tasks and the breaks between them in which the patients were free to behave naturally. During the two days of home monitoring, accelerometer data was recorded continuously at 50Hz, 24 hours a day.     

\section{Modeling} 
\label{sec:ftr-extract}

\subsection{Wavelet features extraction}
\label{sec:wav_ftr-extract}
The wavelet transform  has proven to be an effective tool in signal processing, classification, and clustering \cite{antoniadis2013clustering, hope2015clustering}. Wavelets can handle non-stationary signals better than the Fourier transform, and capture localized information on the time-frequency plane. This property is very important in our case, since we are trying to capture events in time and frequency such as temporary tremor. The wavelet coefficients are obtained by translating and scaling a unique mother wavelet function. For a signal of length $2^k$, we have in total $k$ scales, where each scale represent a different level of frequencies, with higher scales relating to the lower frequency.   
Using signals of a length which is a power of 2 allows us to make use of the  pyramidal algorithm  \cite{mallat1989theory} to obtain the wavelet coefficients at different scales.

Following \cite{antoniadis2013clustering}, we use the absolute and relative contribution of each scale. Specifically, denoting the contribution to the total energy, of each scale $cont_j$,
we then use the relative energy:

\begin{equation}
\label{eq:rel-energy}
rel_j = cont_j/\sum_i(cont_i)
\end{equation}

\noindent For a signal of length $2^k$ we now have in total $2k$ features, k contribution features and k relative energy features. These features have the advantage of being very easy to compute, helpful in understanding the dominant frequencies in our signal, and as we show, very useful for PD symptoms detection.

\subsection{Symptoms estimation}
\begin{algorithm}[H] 

\caption{Rest Tremor Classification}

\begin{algorithmic}[1]
\label{alg:trem}

\REQUIRE Raw signal, tremor labels, tremor thresholds $\theta_1$ and $\theta_2$

\ENSURE tremor prediction $t \in \{0,1,2\}$

\STATE{ $\forall i: $ Extract the features $rel^{X}_{i}$, $rel^{Y}_{i}$ and $rel^{Z}_{i}$ (eq. \ref{eq:rel-energy})}
    \STATE{$\forall i: rel^{(avg)}_{i} \leftarrow (rel^{X}_{i}+rel^{Y}_{i}+rel^{Z}_{i})/3$}
    \STATE{Train a one-vs-all multi-class SVM using the set of  $rel^{(avg)}$ features, with the tremor level target.}
    \STATE{For each axis, predict tremor level: $ pred \in \{0,1,2\}$, using the axis-specific relative features (1) }
    \STATE{$pred^{total} \leftarrow pred^{X}+pred^{Y}+pred^{Z}$}
    \IF {$pred^{total}>\theta_1$}
    	\STATE{predict 2}
    \ELSIF{$pred^{total}>\theta_2$}
    	\STATE{predict 1}
    \ELSE
    	\STATE{predict 0}
    \ENDIF
    
\end{algorithmic}

\end{algorithm}

For each axis (x, y, z) both relative energy features (equation \ref{eq:rel-energy}) and mean relative energy (average over the 3 axes, of the former) are computed for each of the wavelet scales. During training, the mean relative energy features are used in a Support Vector machine (SVM) to predict the tremor level (0, 1, or 2) as annotated by the physician during the task. 

Next, the model is utilized to evaluate a per-axis tremor score by using the  axis-specific relative energy features in the trained SVM. Finally, the sum of the per-axis scores is threshold-ed heuristically to obtain the final prediction (see Algorithm \ref{alg:trem}). The rational behind this approach is that while the tremor manifests often in only a single axis, the physician provides a single score which is then assigned to the data from all three axes.  The final stage may be improved by using a more rigorous  machine learning approach, and learning a function from the per-axis scores to the final classification. 

The model described above in Algorithm \ref{alg:trem}  is used for the  prediction of tremor in resting conditions. The same method is used to predict bradykinesia in gross upper limbs movements, and dyskinesia in rest (both are binary classifications and thus the final decision in rows 6-11 is replaced with a single threshold). We omit the full description of these models due to space considerations.  

\section{Experiments and results}

\subsection{Data processing}
\label{sec:data-processing}

Raw accelerometer data was split into 10 second intervals (500 samples), with 50\% (5 second) overlap between adjacent segments to ensure that momentary patterns to be detected are not missed because of boundary effects. Since the duration of many of the motor tasks is longer than 10 seconds, each interval was assigned the scores associated with the task segment it belonged to. The 10 second intervals were then further processed using a wavelet transform, as described in Section 3.1.

In order to obtain a signal windows of a length which is a power of 2, we use a spline interpolation (in the output we have is a signal with the length $2^k$ where k is the smallest integer such that the length of the signal is smaller than $2^k$). We then apply the Debuchy wavelet transform \cite{daubechies1988orthonormal} on the output and use the relative contribution features as described in Section \ref{sec:wav_ftr-extract}.

\subsection{Evaluation}
All models were evaluated using leave one patient out cross-validation; a 19-fold cross validation, where each group is the the set of intervals which belong to one of the 19 patients. This procedure helps avoid over-fitting our models to specific patients, allowing the models to have global value, and good generalization properties for new patients. 

\begin{figure}[!t] 
\centering
\includegraphics[width=\linewidth, trim={2.7cm 4.2cm 0.5 2.5cm},clip]{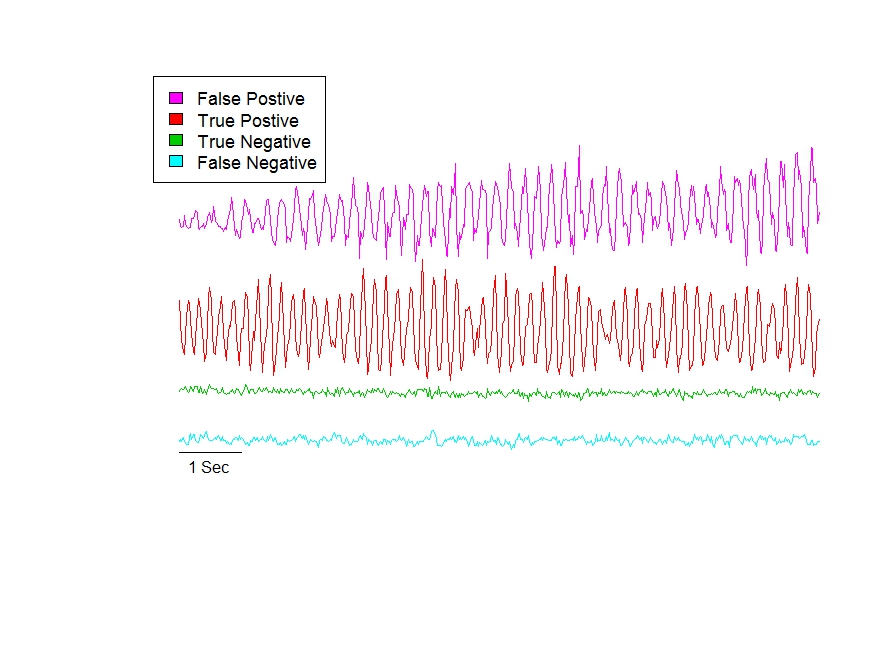}
\caption{Misclassification or mislabeling? Raw accelerometer signals from patients at rest with and without tremor. Visually similar signals were labeled differently by the physician.}
\label{fig:tremor}
\end{figure}

\begin{figure}[!t] 
\centering
\includegraphics[width=\linewidth, trim={2.5cm 4cm 0 2.5cm},clip]{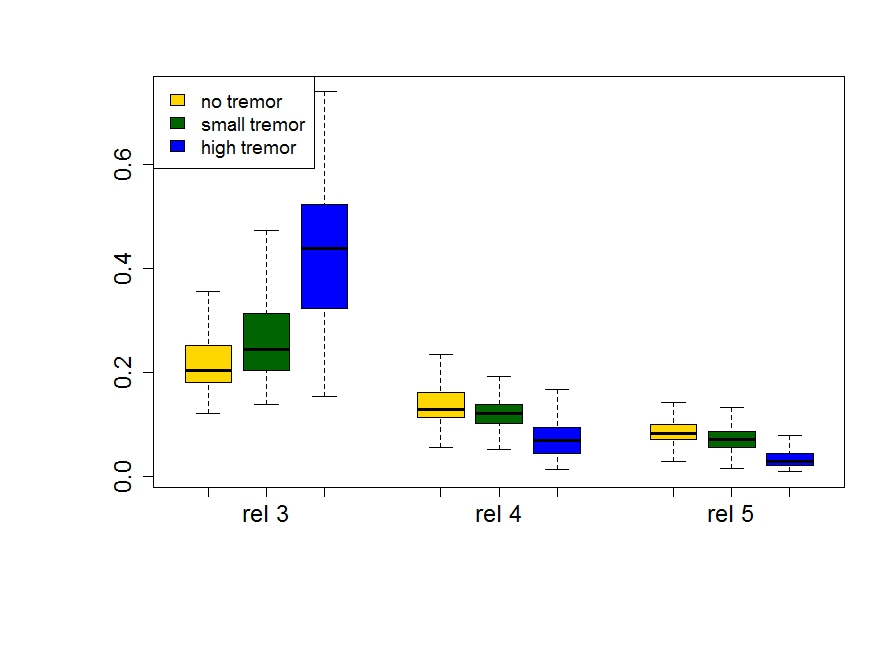}
\caption{Boxplots of the 3, 4, and 5 relative scales, for each tremor level, as assigned by a physician.}
\label{fig:box-trem}
\end{figure}

\subsection{Tremor}
Tremor is among the more visible symptoms of PD, and arguably the most associated with it. Figure \ref{fig:tremor} shows typical signals from a patient at rest labeled by a physician as with or without tremor (second and third lines respectively).

Rest tremor is characterized by a fast oscillation of the accelerometer signal, making the relative energy features most appropriate for classification of the extent of tremor (0, 1, or 2 labels correspond to no-tremor, some-tremor and severe-tremor). Figure \ref{fig:box-trem} shows that the higher frequencies such as $rel_3$ are much more pronounced in strong tremor, than the lower frequencies such as $rel_4, rel_5$. 

The results of our classification model for rest tremor are displayed in Table \ref{table:classification}a. The main source of error is in separating no tremor at all (0 label) and very small tremor such as finger tremor or tremor that is suppressed by the patient (1 label), in both cases the tremor is apparently not detected by the sensor. Moreover, some of our algorithm's mistakes are probably labeling mistakes (see Fig. \ref{fig:tremor} for representative examples).  

We stress that the suggested method is very convenient for use in home-monitoring environments. The data required is derived solely from a wrist-worn smart-watch which does not overly burden the patient, and the model uses fast and standard calculations suitable for real-time systems on any platform. 

\subsection{Dyskinesia}
\subsubsection{Rest dyskinesia}
Dyskinesia is a motor problem involving involuntary movements, and is a known side-effect of long term PD levodopa therapy, affecting up to 80\% of chronically treated patients \cite{nutt1990levodopa}. During rest, we expect these involuntary movements of the body to manifest in a significantly lower frequency band than tremor. 

The binary classifier for this task (Dyskinesia present or not) utilizes both absolute and relative energy features. Data from signal segments where the patient is either sitting or standing are used. We modify the algorithm for tremor classification (Algorithm \ref{alg:trem}) first by using the two kinds of features, and furthermore, since this is now a binary classification we change the thresholding (lines 6-12) to reflect this. A segment is classified as positive if at least one of the axes received a positive label. An ROC curve was computed by averaging the per-axis probability derived from an SVM, leading to an AUC of 75\%.

\subsubsection{Walking dyskinesia}
While we can detect dyskinesia in rest, the task of monitoring strong dyskinesia in active individuals is substantially more complicated since intentional and involuntary movements overlap both in frequency bands and in structure. Yet, for a known well-structured activity, this is still possible. 

During regular walking, the majority of the accelerometer signal is limited to a single axis (we denote this the X axis). We compute for each energy level the relative contribution of the X axis, relative to the other two axes (see section \ref{sec:data-processing} for the definition of the $cont^{axis}_i$ features) as follows:

\[ 
\forall i: w_i = \frac{cont^y_i cont^z_i}{cont^x_i} 
\]

Here again we use an SVM with 19-fold cross validation. The AUC in a leave one-patient out cross-validation obtained for this classification  is $0.92$ (see confusion matrix in Table \ref{table:classification}c).

\subsection{Bradykinesia}
Much work has been done on detecting bradykinesia from hand rotation and other periodic hand movements \cite{jun2011quantification}. Here we attempt to monitor bradykinesia in gross upper-limb movements such as folding towels, organizing sheets of paper in a folder, and reaching out to take a glass of water and drinking it. These more natural behaviors are intended to approximate at-home monitoring of patients during everyday activities.
In this task, we again use the relative features in the classification process. The AUC for the ROC obtained is 70\%.  

\begin{table}[t!]
\centering
\caption{Confusion matrices for classification of tremor, dyskinesia, and bradykinesia.}
\subfloat[resting tremor]
{\begin{tabular}{|c|c|c|c|}
	   \hline
       True/Pred& 0 & 1 & 2 \\
       \hline \hline
      0&1531&28&8\\
      \hline
      1&390&159&10\\
      \hline
      2&22&52&218\\
      \hline
     \end{tabular}
}
\hfill
\subfloat[resting dyskinesia]
{\begin{tabular}{|c|c|c|c|} 
\hline True/Pred&0&1\\ \hline \hline 0&1264&4\\ \hline 1&150&123\\ \hline
\end{tabular}
}
\hfill
\subfloat[walking dyskinesia]
{\begin{tabular}{|c|c|c|c|} 
\hline True/Pred &0&1\\ \hline \hline 0&2244&19\\ \hline 1&145&309\\ \hline
\end{tabular}
}
\\
\subfloat[gross upper limb bradykinesia]
{\begin{tabular}{|c|c|c|c|} 
\hline True/Pred &0&1\\ \hline \hline 0&2177&201\\ \hline 1&340&311\\ \hline
\end{tabular}
}
\label{table:classification}
\end{table}

\section{Conclusion}
\label{sec:conclusion}

PD is a prevalent neurodegenerative disease, which manifests predominantly as a movement disorder. As such, it is an interesting use-case for the introduction of wearable devices and sensors as a means of cheap and continuous monitoring of patient condition.

We present here a novel wavelet-based method for calculating objective measures related to the major Parkinsonian symptoms, and monitoring targets: tremor, dyskinesia and bradykinesia. Whereas previous work focused on handcrafting specific features for classification of the various symptoms of interest, the approach adopted here allows the same simple and systematic features to be applied in the models pertaining to all of the symptoms. We show high performance in terms of detecting symptoms and severity using a single wrist-worn smart-watch. 

The central limitation of the current approach is that a single wrist-mounted sensor will inherently miss symptom related movements which occur solely in other parts of the body. We intend to overcome this issue in the future by aggregating symptom predictions over time, thus in effect waiting long enough for the characteristic movement to appear in the limb we are able to measure from. 

An important goal of this research is to inspire wearable technology adoption  outside of the laboratory setting. Indeed, our results demonstrate that resting  tremor can be detected at relatively high accuracy, a finding that likely translates well to the home environment. Future work will additionally focus on evaluation of increasingly native environments towards achieving long term monitoring of PD patients in their daily lives.  

\section{Acknowledgments}
This work is part of the collaboration between the Advanced Analytics health team at Intel and the Michael J. Fox Foundation for Parkinson’s disease research. We thank Zeev Waks and Tom Hope for insightful discussions about this work.  

\bibliographystyle{IEEEbib}
\bibliography{lib}

\end{document}